\documentclass[15pt]{article}
\usepackage[super,compress]{cite}
\usepackage{graphicx}
\usepackage{caption,subcaption}
\usepackage{url,hyperref}
\usepackage{color}
\usepackage{amsmath,amsfonts,amssymb}

\def\beq{\begin{equation}}
\def\eeq{\end{equation}}

\date{} 
\begin{document}

\title{ The ANAIS Dark Matter Project: Status and Prospects}
\maketitle
\begin{center}{Julio Amar\'e, Susana Cebri\'an, Clara Cuesta\footnote{Present address: Center for Experimental Nuclear Physics and Astrophysics and Department of Physics, University of Washington, WA, US}, 
Eduardo Garc\'ia,
Mar\'ia Mart\'inez$^\star$\footnote{Present address: Universit\`a degli Studi di Roma "La Sapienza", Roma, Italy}, 
Miguel \'Angel Oliv\'an, Ysrael Ortigoza, Afonso Ortiz de Sol\'orzano, 
Carlos Pobes\footnote{Present address: Instituto de Ciencia de Materiales de Arag\'on, Universidad de Zaragoza - CSIC, Zaragoza, Spain}, 
Jorge Puimed\'on, Mar\'ia Luisa Sarsa, Jos\'e \'Angel Villar, Patricia Villar}

\vspace{0.5cm}

Laboratorio de F\'isica Nuclear y Astropart\'iculas, Universidad de Zaragoza, \\
C/ Pedro Cerbuna 12, 50009 Zaragoza, SPAIN\\
Laboratorio Subterr\'aneo de Canfranc, Paseo de los Ayerbe s.n., 22880 Canfranc Estaci\'on,
Huesca, SPAIN\\
$^\star$E-mail: mariam@unizar.es\\
\end{center}

\begin{abstract}
The ANAIS (Annual modulation with NaI(Tl) Scintillators) experiment aims at the confirmation of the DAMA/LIBRA
positive annual modulation signal using the same target and technique at the Canfranc Underground Laboratory (LSC).
A first step, named ANAIS--25 (two 12.5~kg NaI(Tl) modules) taking data from December 2012 to February 2015, provided
interesting outcomes: very high light collection efficiency, that could allow to lower the analysis energy threshold down to
the level of 1 keVee, and a good understanding of the different background components, 
in particular the cosmogenic activated isotopes in the crystal bulk and other radioactive contaminations of the NaI crystal/powder.
 But those prototypes clearly pointed to the need for improved crystal radiopurity, in particular for $^{210}$Pb contamination.
Since then, 
improvements in the purification and growing procedures in order to reduce background in the very low energy region have been implemented
and a new 12.5~kg module has been constructed
and installed between the former two crystals, forming the ANAIS--37 setup.
 Very preliminary results of this setup evidence the improvement on radiopurity of the new crystal and are presented here. 
In addition, background simulations and prospects for the full experiment are discussed.
\end{abstract}

\section{Introduction}
The ANAIS experiment aims
to a model independent confirmation of the positive Dark Matter (DM) signal reported by the 
DAMA/NaI\cite{DAMA-NaI} and DAMA/LIBRA\cite{LIBRA} experiments, that have observed an annually modulated
event rate in the very low energy range (from 2 - 6 keVee), as expected from a DM particle
component in the galactic halo, and  whose characteristics are 
very hard to mimic by known backgrounds.
In order to confirm this signal, ANAIS will install $\sim$100~kg of NaI(Tl) scintillator detectors 
in the Canfranc Underground Laboratory (LSC, Spain). 
This paper starts with a brief description of the ANAIS prototypes that have served to 
characterize and improve the detectors in the past years (section \ref{sec:prototypes}), 
and then a review of the detectors performance achieved so far (section \ref{sec:performance}).
Finally, in section \ref{sec:prospects} the experiment status and its prospects in terms of 
sensitivity to annual modulation are discussed.

\section{ANAIS Prototypes}
\label{sec:prototypes}
In an extensive effort to achieve the best detector performance, several prototypes have been tested at LSC in the last years.
ANAIS--0\cite{ANAIS0}, a single 9.6~kg Saint-Gobain\footnote{Saint-Gobain, http://www.saint-gobain.com/} crystal, 
allowed for a valuable understanding of the different background contributions\cite{bkgModel,K40},
estimate of NaI(Tl) slow scintillation constants\cite{scintQuartz, scint}, 
and the establishment of the basis for the bulk event data selection protocols used in the subsequent 
experiments\cite{eventSelection}.
The ANAIS--25 setup\cite{ANAIS25}, operated from December 2012 until March 2015, 
consisted in two 12.5~kg cylindrical (4.75''~$\phi~\times~$11.75'' length) NaI(Tl) crystals  made by 
Alpha Spectra (AS)\footnote{Alpha Spectra Inc., Grand Junction, Colorado, US. http://www.alphaspectra.com/}, 
and labeled as D0 and D1 in the following (see Fig.~\ref{fig:ANAIS37} left).
Each module was coupled to two photomultipliers (PMTs)
through quartz windows, without light guides to improve light collection and was equipped with 
a Mylar window in the lateral face of the copper encapsulation. This window allows to calibrate the detectors
in the very low energy region with external sources 
($^{57}$Co and $^{109}$Cd), that are mounted along flexible wires and introduced into the
shielding through a closed Rn-free tube.
As it will be shown in the next section, the modules design proved to be very efficient in terms of light collection, 
and has been adopted for the final experiment.
The detectors were enclosed in a 30~cm lead shielding plus anti-radon box and active vetoes.
The same setup was used to test a new module produced by AS with improved protocols in order to reduce 
internal contaminants. The new module, labeled as D2, was installed at LSC between the ANAIS--25 ones, in a new setup 
called ANAIS--37 (see Fig.~\ref{fig:ANAIS37} right). 
As concerns the photomultipliers, Hamamatsu R12669SEL2 have been selected for their low background characteristics and
high quantum efficiency ($>$32\%).

\def\figsubcap#1{\par\noindent\centering\footnotesize(#1)}
\begin{figure}[h]
\begin{center}
\parbox{2.1in}{\includegraphics[width=2in]{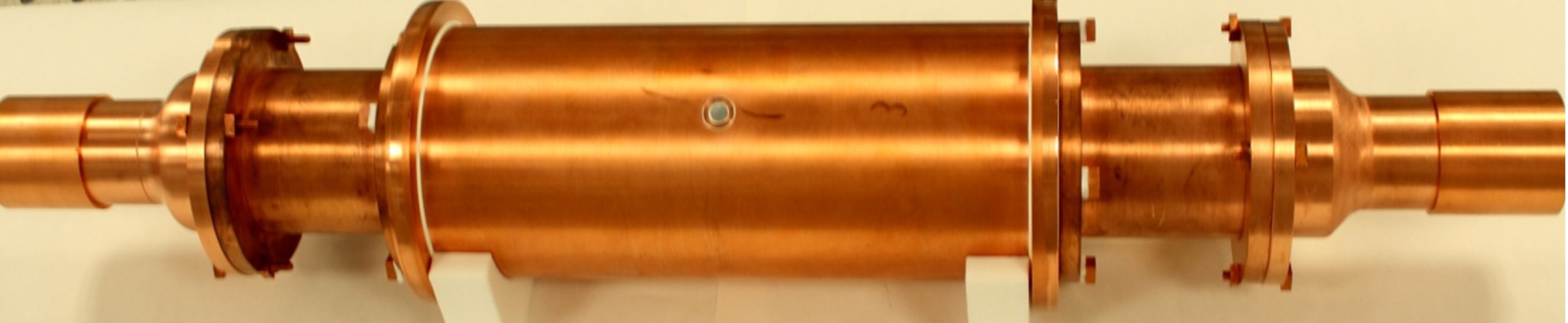}\figsubcap{a}}
\hspace*{4pt}
\parbox{2.1in}{\includegraphics[width=2in]{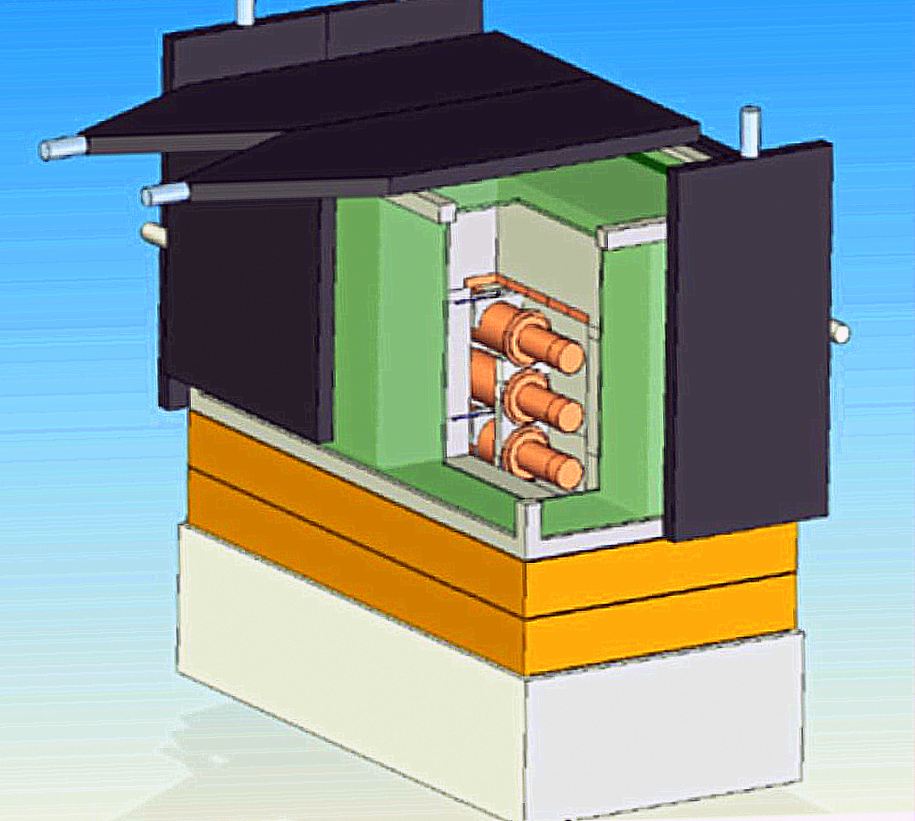}\figsubcap{b}}
\end{center}
\caption{(a) One of the ANAIS--25 modules. The Mylar window for low energy calibration is visible in the center of the copper housing. (b) Sketch of the ANAIS--37 steup. The new module D2, with improved radiopurity, was installed between the ANAIS--25 ones (D0 and D1). }
\label{fig:ANAIS37}
\end{figure}
\section{Detectors Performance}
\label{sec:performance}
\subsection{Light collection and energy threshold}
We determine the light collection in ANAIS modules by a fully characterization of the PMTs single electron response (SER)\cite{ser}.
Then, the number of photoelectrons (phe) collected per energy unit is obtained by 
dividing the mean value of the charge distribution corresponding to a known energy deposition in the detector
(usually 22.6~keV from a calibration with $^{109}$Cd source) by the mean value of the SER charge distribution.
We can report an excellent light collection efficiency in all the AS modules characterized with this method:
around 15 phe/keV (see first column in Tab.~\ref{tab:internalCont}), that represents twice the value obtained for the ANAIS--0 prototype. 
This result points to a lower energy threshold in the AS modules with respect to that of the ANAIS--0 prototype, 
set to 2~keVee\cite{eventSelection}. In fact the 0.9~keV energy deposition following EC decay in $^{22}$Na (an internal contaminant of cosmogenic origin) 
is clearly visible in the ANAIS--37 detectors
energy spectra when the corresponding line from the daughter nucleus de-excitation (1274.5~keV) is detected in another module. 
We are still working in the evaluation of the efficiency at this very low energy 
line, because the event rate in this region is greatly dominated by noise events whose origin is not bulk scintillation\cite{eventSelection}.
\begin{table}[h]
\begin{center}
{\begin{tabular}{@{}cccc@{}}
\hline
Detector & Light Collected  & $^{40}$K     &  $^{210}$Pb  \\
 & phe/keV & mBq/kg & mBq/kg \\
\hline
ANAIS--0 & 7.38$\pm$0.07 & 12.7$\pm$0.6& 0.188$\pm$0.005 \\
D0      & 15.6$\pm$0.2  & 1.4$\pm$0.2 & 3.15$\pm$0.1 \\
D1      & 15.2$\pm$0.1  & 1.1$\pm$0.2 & 3.15$\pm$0.1 \\
D2      & 16.3$\pm$0.6  & 1.1$\pm$0.2 & 0.70$\pm$0.1 \\
 
\hline
\end{tabular}
}
\caption{Light collected and main internal contaminations measured in ANAIS prototypes. }
\label{tab:internalCont}
\end{center}
\end{table}

\subsection{Background}
\begin{figure}[h]
\begin{center}
\parbox{2.4in}{\includegraphics[width=2.5in]{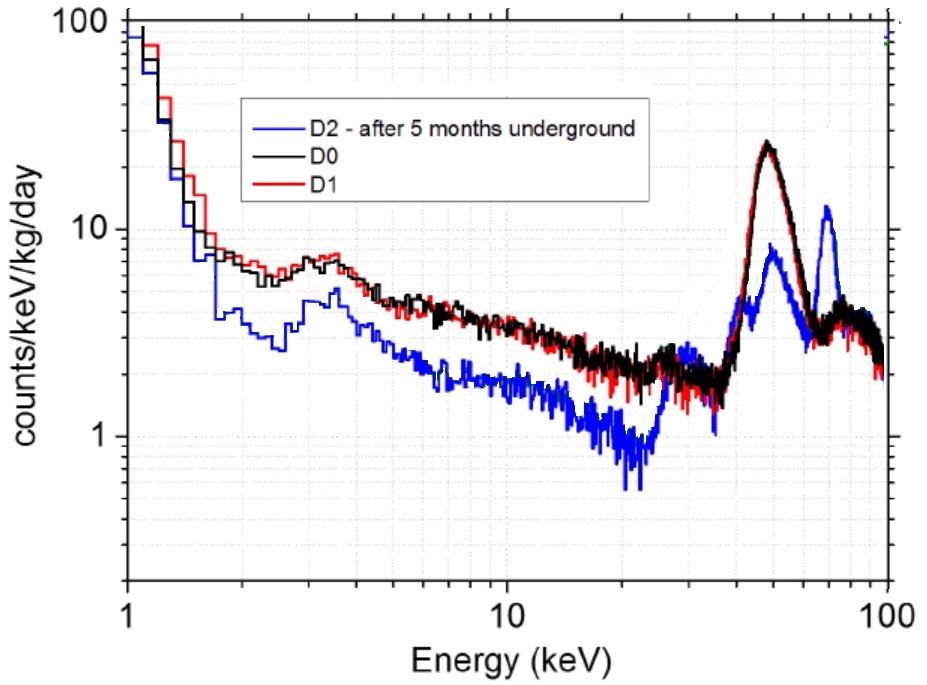}\figsubcap{a}}
\hspace*{0pt}
\parbox{2.2in}{\includegraphics[width=2.3in]{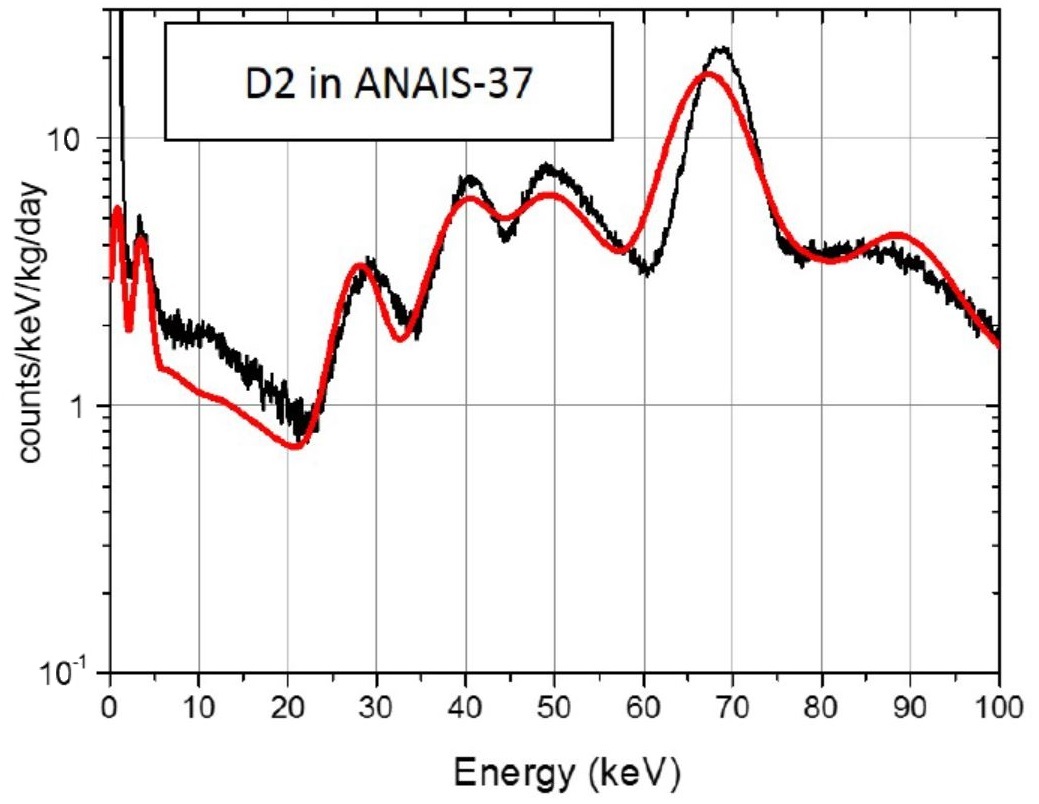}\figsubcap{b}}
\end{center}
\caption{a) Background levels measured for the three modules of ANAIS--37 setup. 
Cosmogenically activated isotopes are still noticeable in D2 after 5 months underground.
b) Comparison between the experimental background measured by D2 prototype (black) and our still under construction background 
model based on Ref.~\citenum{bkgModel} (red).}
\label{fig:bkg}
\end{figure}

All the external components (copper housing, quartz windows, PMTs, near electronics, shielding,..) have been screened for 
radiopurity at LSC by HPGe spectroscopy, obtaining only upper limits in most cases. The most limiting elements are the PMTs,
but our Monte Carlo simulations show that the resulting background below 10~keVee is at 0.1~cpd/keV/kg level,
being the contamination of the detector itself 
the dominant contribution\cite{bkgModel}.
Both ANAIS--25 and ANAIS--37 setups started taking data very soon after their installation underground, therefore 
they produced valuable information regarding cosmogenic isotopes activated in the NaI powder/crystal during the 
crystal growing and mounting stages\cite{Cosmogenics}. Most of those 
isotopes are short-living and not dangerous for the experiment, except $^{22}$Na and $^3$H.
Apart from that, the most important internal contaminants are $^{40}$K and $^{210}$Pb. The former can be determined by studying the 
coincidence between a 3.2~keV energy deposition in one crystal 
and 1460.8~keV on another one. The latter can be identified by alpha spectroscopy, profiting from the different scintillation constant
of alpha and beta/gamma particles. The last two columns of Tab.~\ref{tab:internalCont} show the activity identified for these two
isotopes for the three AS modules in comparison with that of ANAIS--0. 
A clear reduction in $^{40}$K contamination can be reported, with a present level of $\sim$1~mBq/kg corresponding to 
around 40~ppb natural potassium, but a higher level of $^{210}$Pb content with respect to the Saint Gobain
crystals was noticed in the first two modules. After a careful revision of the crystal growing and handling protocols 
the problem was partially solved in the D2 detector. 
Further reduction in the next modules is being pursued, collaborating with AS in the improvement of all the purification/growing/handling protocols.
The corresponding reduction in the low energy background level of D2 with respect to D0 and D1 is evident in Fig.~\ref{fig:bkg} left.
Successful background models have been developed for the ANAIS--0 
and ANAIS--25 setups down to 3 keV\cite{bkgModel,bkgModelA25}. 
The background model of D2 in ANAIS--37 is still under construction. 
As can be checked in Fig.~\ref{fig:bkg} right, 
the measured background is well accounted for above 20~keV 
but some background contributions at low energy should still 
be properly identified.  
Work is in progress in order to better understand this background. 

\section{Status and Prospects}
\label{sec:prospects}
\begin{figure}[h]
\begin{center}
\parbox{2.2in}{\includegraphics[width=2.4in]{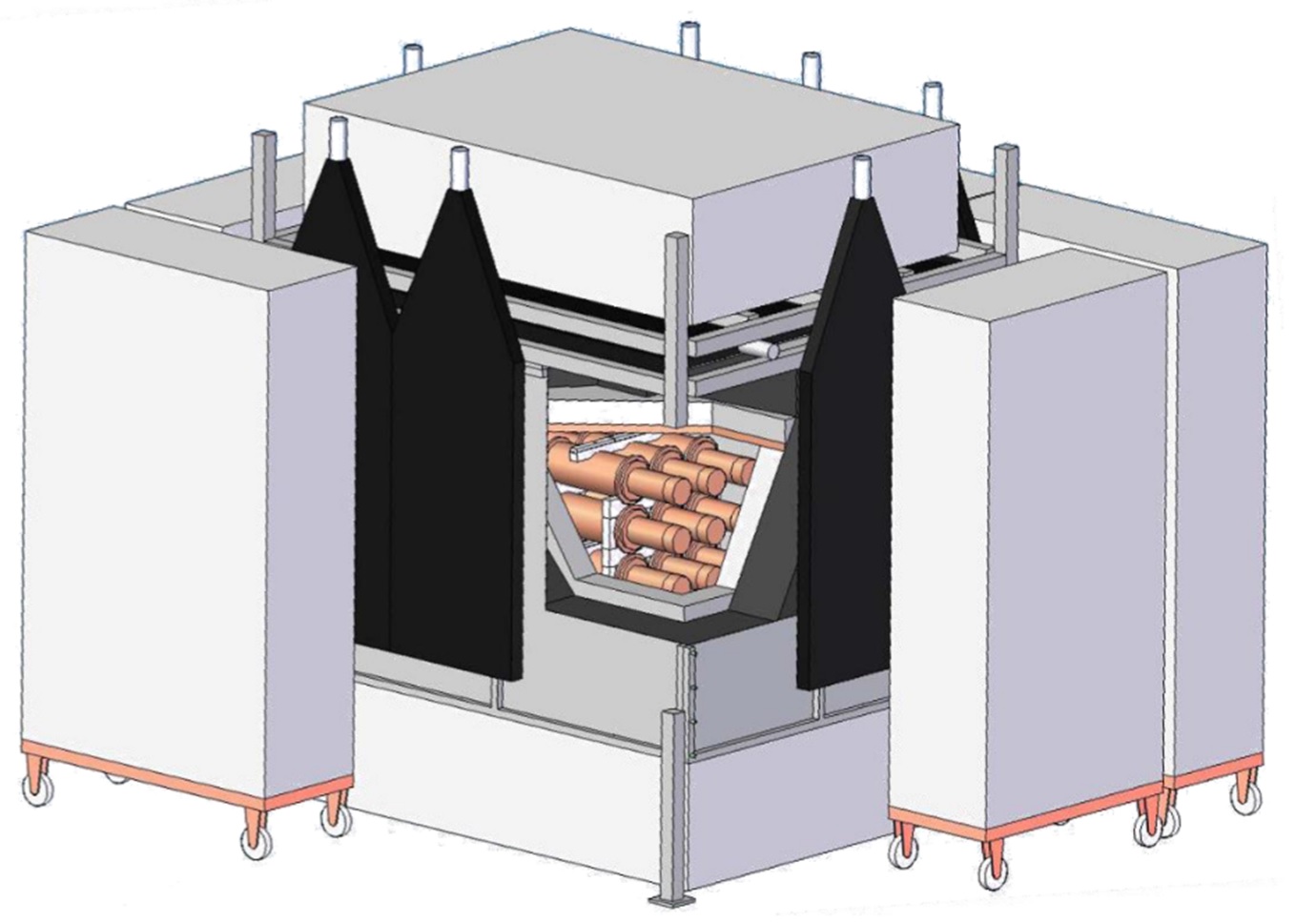}\figsubcap{a}}
\hspace*{2pt}
\parbox{2.2in}{\includegraphics[width=2.4in]{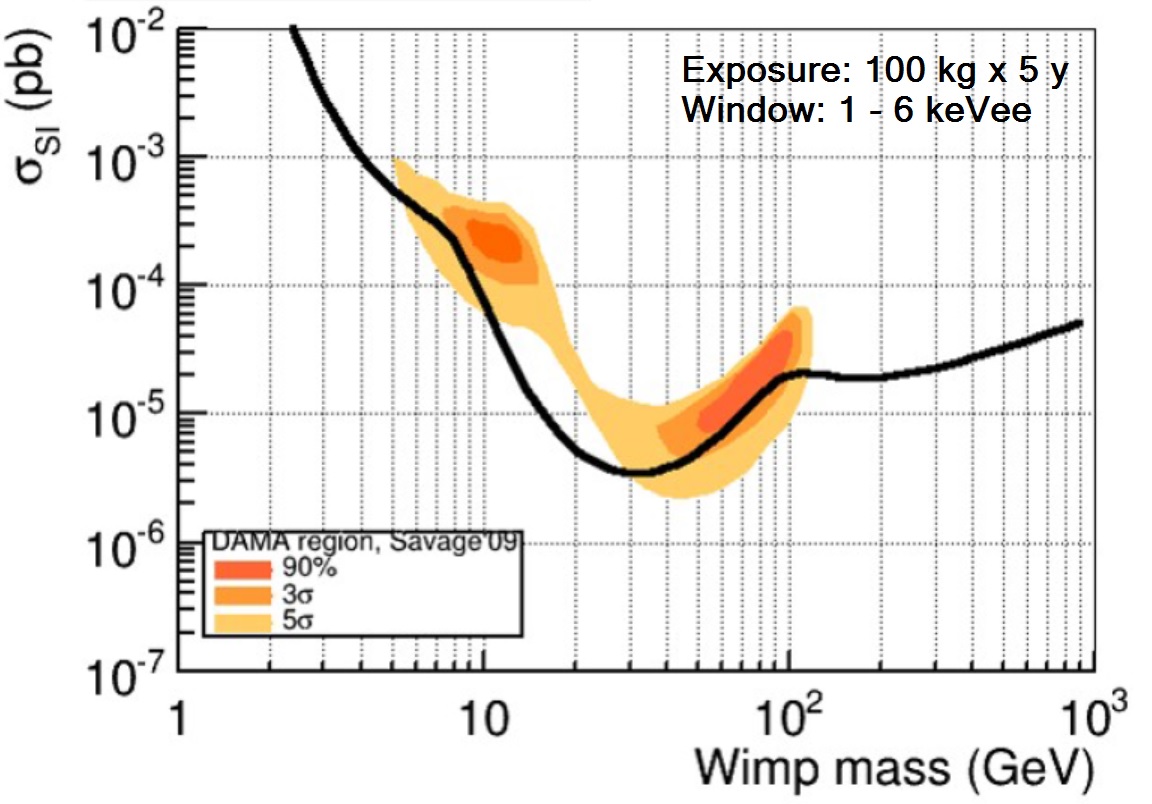}\figsubcap{b}}
\end{center}
\caption{a) Design of the future 3x3 ANAIS matrix inside the complete shielding. b) Expected sensitivity of ANAIS in a conservative scenario (see text for details).}
\label{fig:sensitivity}
\end{figure}

A new more radiopure AS module is expected to arrive at LSC in January 2016, and 5 more
modules will arrive along 2016, to start taking data with 112.5~kg NaI(Tl) in 
a 3$\times$3 matrix configuration by the second half of 2016.
A sketch of the full experiment is shown if Fig.~\ref{fig:sensitivity} left. 
The shielding will consist of 10~cm of archaeological lead,
20~cm of low activity lead, 40 cm of neutron moderator, an anti-radon box, and an active muon
veto system made up of plastic scintillators covering top and sides of the whole setup.
Active veto system, data acquisition and analysis software 
and slow control system are fully commissioned at LSC.

The $^{22}$Na and $^{40}$K contributions in the low energy region can be substantially reduced tagging those 
events by the detection of the high energy gammas emitted simultaneously. 
The background reduction power of the 3x3 matrix working in anticoincidence can be 
considerably increased by surrounding the detectors with a liquid scintillator functioning as active veto.
We are working in the evaluation of the possibilities of such a system and its feasibility
in the context of the ANAIS project.

Finally, in Fig.~\ref{fig:sensitivity} right we present the projected sensitivity (spin independent) 
to DM annual modulation of the experiment in the
conservative scenario described above, supposing 100~kg of
NaI(Tl), 5 years of data taking and the background already achieved.
In the calculation, we have considered Helm form factors, a standard halo
model (isothermal sphere) with $\rho$=0.3~GeV/cm$^3$, $v_0$=220~km/s and $v_{esc}$=650~km/s,
quenching factors 0.3 for Na and 0.1 for I, and an energy window to
look for the modulated signal from 1 to 6~keVee.
Even in this conservative scenario,  a 90\% C.L. positive signal would be obtained in 
90\% of the carried out experiments for most of the DAMA/LIBRA singled-out
DM parameter space.

In summary, good quality NaI(Tl) detectors from AS have been fully characterized, showing an
outstanding light collection and 1~keVee energy threshold at reach.
The potassium content is around 40 ppb, and $^{210}$Pb bulk contamination 
is at 0.7 mBq/kg level, that means a factor 5 improvement with respect to 
previous prototypes. Further improvement in both isotopes is expected in next modules.
ANAIS foresees to start taking data with 112.5~kg along 2016, with good sensitivity prospects 
for exploring the DAMA/LIBRA signal even under very conservative assumptions.

\section*{Acknowledgments}
This work has been supported by the Spanish Ministerio de Econom\'ia y Competitividad and
the European Regional Development Fund (MINECO-FEDER) (FPA2011-23749 and FPA2014-
55986-P), the Consolider-Ingenio 2010 Programme under grants MULTIDARK CSD2009- 00064
and CPAN CSD2007-00042, Gobierno de Arag\'on and the European Social Fund (Group
in Nuclear and Astroparticle Physics). P. Villar is supported by the MINECO Subprograma de
Formaci\'on de Personal Investigador. We also acknowledge LSC and GIFNA staff for their
support.

\end{document}